\documentclass[a4paper,british,numberwithinsect]{lipics}

\usepackage{braket,stmaryrd}
\usepackage{ifthen}
\usepackage{tikz}
\usetikzlibrary{arrows}

\usepackage[pagewise,mathlines,switch]{lineno}

\newcommand*\patchAmsMathEnvironmentForLineno[1]{%
  \expandafter\let\csname old#1\expandafter\endcsname\csname #1\endcsname
  \expandafter\let\csname oldend#1\expandafter\endcsname\csname end#1\endcsname
  \renewenvironment{#1}%
     {\linenomath\csname old#1\endcsname}%
     {\csname oldend#1\endcsname\endlinenomath}}%
\newcommand*\patchBothAmsMathEnvironmentsForLineno[1]{%
  \patchAmsMathEnvironmentForLineno{#1}%
  \patchAmsMathEnvironmentForLineno{#1*}}%
\AtBeginDocument{%
\patchBothAmsMathEnvironmentsForLineno{equation}%
\patchBothAmsMathEnvironmentsForLineno{align}%
\patchBothAmsMathEnvironmentsForLineno{flalign}%
\patchBothAmsMathEnvironmentsForLineno{alignat}%
\patchBothAmsMathEnvironmentsForLineno{gather}%
\patchBothAmsMathEnvironmentsForLineno{multline}%
}

\newcommand{\struct}[1]{\mathfrak{#1}}
\newcommand{\class}[1]{\mathcal{#1}}
\newcommand{\aut}[1]{\mathcal{#1}}

\newcommand{\univ}[1]{\Vert #1\Vert}
\newcommand{\ar}{\operatorname{ar}}
\newcommand{\dom}{\operatorname{dom}}

\newcommand{\tp}{\operatorname{tp}}

\newcommand{\Rel}{\mathcal{R}}
\newcommand{\Nat}{\mathbb{N}}
\newcommand{\restrict}[1]{\mathord{\restriction} #1}
\newcommand{\run}[3]{#1=#2,\dotsc,#3}
\newcommand{\elem}[3]{#1\in\{#2,\dotsc,#3\}}
\newcommand{\trees}[2][\empty]{T_{#2\ifthenelse{\equal{#1}{\empty}}{}{,#1}}}
\newcommand{\VD}{\class{VD}}
\newcommand{\rankVD}{\operatorname{VD}}
\newcommand{\rankFC}{\operatorname{FC}}
\newcommand{\rankCB}{\operatorname{CB}}

\newcommand{\Bits}{\{0,1\}}
\newcommand{\FO}{\mathsf{FO}}
\newcommand{\FOinf}{\FO(\exists^\infty)}
\newcommand{\Tbin}[1][\empty]{\struct T_{2\ifthenelse{\equal{#1}{\empty}}{}{,#1}}}
\newcommand{\Strings}{\Bits^\star}

\theoremstyle{plain}
\newtheorem{proposition}[theorem]{Proposition}

\title{The Rank of Tree-Automatic Linear Orderings}
\author{Martin Huschenbett}
\affil{Institut f\"ur Theoretische Informatik, Technische Universit\"at Ilmenau, Germany\\\href{mailto:martin.huschenbett@tu-ilmenau.de}{\nolinkurl{martin.huschenbett@tu-ilmenau.de}}}

\begin{document}


\maketitle

\thispagestyle{empty}

\begin{abstract}
We generalise Delhomm\'e's result that each tree-automatic ordinal is strictly below $\omega^{\omega^\omega}$ by showing that any tree-automatic linear ordering has $\rankFC$-rank strictly below $\omega^\omega$. We further investigate a restricted form of tree-automaticity and prove that every linear ordering which admits a tree-automatic presentation of branching complexity at most $k\in\Nat$ has $\rankFC$-rank strictly below $\omega^k$.
\end{abstract}


\section{Introduction}

In \cite{Del04}, Delhomm\'e showed that an ordinal $\alpha$ is string-automatic if, and only if, $\alpha<\omega^\omega$ and it is tree-automatic if, and only if, $\alpha<\omega^{\omega^\omega}$. Khoussainov, Rubin, and Stephan \cite{KRS05} extended his technique to prove that every string-automatic linear ordering has finite $\rankFC$\nobreakdash-rank. Although it is commonly expected that every tree-automatic linear ordering has $\rankFC$-rank below $\omega^\omega$, this conjecture has not been verified yet.\footnote{Recently, Jain, Khoussainov, Schlicht, and Stephan \cite{JKS12} independently from us obtained results which verify this conjecture as well.} We close this gap by providing the missing proof (Theorem~\ref{thm:main}). As part of this, we give a full proof of Delhomm\'e's decomposition theorem for tree-automatic structures (Theorem~\ref{thm:delhomme}). Afterwards, we investigate a restricted form of tree-automaticity where the branching complexity of the trees involved is bounded. We show that each linear ordering which admits a tree-automatic presentation of branching complexity $k\in\Nat$ has $\rankFC$-rank below $\omega^k$ (Theorem~\ref{thm:main_bounded_rank}). As a consequence, we obtain that an ordinal $\alpha$ admits a tree-automatic presentation whose branching complexity is bounded by $k$ if, and only if, $\alpha<\omega^{\omega^k}$.

\section{Tree-Automatic Structures}

This section recalls the basic notions of tree-automatic structures (cf.~\cite{BGR11,Blu99}).

Let $\Sigma$ be an alphabet. The set of all \emph{(finite) words} over $\Sigma$ is denoted by $\Sigma^\star$ and the \emph{empty word} by~$\varepsilon$. A \emph{tree domain} is a finite, prefix-closed subset $D\subseteq\Strings$. The \emph{boundary} of $D$ is the set $\partial D=\set{ ud | u\in D,d\in\Bits,ud\not\in D}$ if $D$ is not empty and $\partial\emptyset=\{\varepsilon\}$ otherwise. A \emph{$\Sigma$-tree} (or just \emph{tree}) is a map $t\colon D\to\Sigma$ where $\dom(t)=D$ is a tree domain. The \emph{empty tree} is the unique $\Sigma$-tree $t$ with $\dom(t)=\emptyset$. The set of all $\Sigma$-trees is denoted by $\trees\Sigma$ and its subsets are called \emph{(tree) languages}. For $t\in\trees\Sigma$ and $u\in\dom(t)$ the \emph{subtree} of $t$ rooted at $u$ is the tree $t\restrict u\in\trees\Sigma$ defined by
\begin{equation*}
	\dom(t\restrict u) = \set{ v\in\{0,1\}^\star | uv\in\dom(t) }
	\quad\text{and}\quad
	(t\restrict u)(v) = t(uv)\,.
\end{equation*}
For $u_1,\dotsc,u_n\in\dom(t)\cup\partial\dom(t)$ which are mutually no prefixes of each other and trees $t_1,\dotsc,t_n\in\trees\Sigma$ we consider the tree $t[u_1/t_1,\dotsc,u_n/t_n]\in\trees\Sigma$, Intuitively, $t[u_1/t_1,\dotsc,u_n/t_n]$ is obtained from $t$ by simultaneously replacing for each $\run i1n$ the subtree rooted a $u_i$ by $t_i$. Formally,
\begin{equation*}
	\dom\bigl(t[u_1/t_1,\dotsc,u_n/t_n]\bigr) =
		\dom(t)\setminus\bigl(\{u_1,\dotsc,u_n\}\{0,1\}^\star\}\bigr)\cup
		\bigcup_{1\leq i\leq n} \{u_i\}\dom(t_i)
\end{equation*}
and
\begin{equation*}
	\bigl(t[u_1/t_1,\dotsc,u_n/t_n]\bigr)(u) = \begin{cases}
		t_i(v) & \text{if $u=u_iv$ for some (unique) $\elem i1n$}\,, \\
		t(u) & \text{otherwise}\,.
	\end{cases}
\end{equation*}

A \emph{(deterministic bottom-up) tree automaton} $\aut A=(Q,\iota,\delta,F)$ over $\Sigma$ consists of a finite set $Q$ of \emph{states}, a \emph{start state} $\iota\in Q$, a \emph{transition function} $\delta\colon \Sigma\times Q\times Q\to Q$, and a set $F\subseteq Q$ of \emph{accepting} states. For all $t\in\trees\Sigma$, $u\in\dom(t)\cup\partial\dom(t)$, and maps $\rho\colon U\to Q$ with $U\subseteq\partial\dom(t)$ a state $\aut A(t,u,\rho)\in Q$ is defined recursively by
\begin{equation*}
	\aut A(t,u,\rho) = \begin{cases}
		\delta\bigl(t(u),\aut A(t,u0,\rho),\aut A(t,u1,\rho)\bigr) & \text{if $u\in\dom(t)$,} \\
		\rho(u) & \text{if $u\in U$,} \\
		\iota & \text{if $u\in\partial\dom(t)\setminus U$.}
	\end{cases}
\end{equation*}
The second parameter is omitted if $u=\varepsilon$ and the third one if $U=\emptyset$. Notice that ${\aut A(t,u)=\aut A(t\restrict u)}$. The tree language \emph{recognised} by $\aut A$ is the set
\begin{equation*}
	L(\aut A) = \Set{ t\in\trees\Sigma | \aut A(t)\in F }
\end{equation*}
of all trees which yield an accepting state at their root. A language $L\subseteq\trees\Sigma$ is \emph{regular} if it can be recognised by some tree automaton.

Let $\Box\not\in\Sigma$ be a new symbol and $\Sigma_\Box=\Sigma\cup\{\Box\}$. The \emph{convolution} of an $n$-tuple ${\bar t=(t_1,\dotsc,t_n)\in(\trees\Sigma)^n}$ of trees is the tree $\otimes\bar t\in\trees{\Sigma_\Box^n}$ defined by
\begin{equation*}
	\dom(\otimes\bar t)=\dom(t_1)\cup\dotsb\cup\dom(t_n)
	\quad\text{and}\quad
	(\otimes\bar t)(u)=\bigl(t'_1(u),\dotsc,t'_n(u)\bigr)\,,
\end{equation*}
where $t'_i(u)=t_i(u)$ if $u\in\dom(t_i)$ and $t'_i(u)=\Box$ otherwise. A relation $R\subseteq(\trees\Sigma)^n$ is \emph{automatic} if the tree language
\begin{equation*}
	\otimes R = \set{ \otimes\bar t | \bar t\in R }\subseteq\trees{\Sigma_\Box^n}
\end{equation*}
is regular. We say a tree automaton \emph{recognises} $R$ if it recognises $\otimes R$.

A \emph{(relational) signature} $\tau=(\Rel,\ar)$ is a finite set $\Rel$ of \emph{relation symbols} together with an \emph{arity map} $\ar\colon\Rel\to\Nat_+$. A $\tau$-structure $\struct A=\bigl(A;(R^{\struct A})_{R\in\Rel}\bigr)$ consists of a set $A=\univ{\struct A}$, its \emph{universe}, and an $\ar(R)$-ary relation $R^{\struct A}\subseteq A^{\ar(R)}$ for each $R\in\Rel$.\footnote{By convention, structures are named in Fraktur and their universes by the same letter in Roman.} Given a subset $B\subseteq A$, the \emph{induced substructure} $\struct A\restrict B$ is defined by
\begin{equation*}
	\univ{\struct A\restrict B} = B
	\quad\text{and}\quad
	R^{\struct A\restrict B} = R^{\struct A}\cap B^{\ar(R)}\ \text{for $R\in\Rel$.}
\end{equation*}

\emph{First order logic} $\FO$ over $\tau$ is defined as usual and $\FOinf$ is its extension by the ``there exist infinitely many''-quantifier $\exists^\infty$. Writing $\phi(x_1,\dotsc,x_n)$ means that all free variables of the formula $\phi$ are among the $x_i$. For a formula $\phi(x_1,\dotsc,x_m,y_1,\dotsc,y_n)$ and a tuple $\bar b\in A^n$ we let
\begin{equation*}
	\phi^{\struct A}\bigl(\cdot,\bar b\bigr) = \Set{ \bar a\in A^m | \struct A\models\phi\bigl(\bar a,\bar b\bigr) }\,.
\end{equation*}
If $n=0$ we simply write $\phi^{\struct A}$ instead of $\phi^{\struct A}(\cdot)$.

\begin{definition}
A \emph{tree-automatic presentation} of a $\tau$-structure $\struct A$ is a tuple $\bigl(\aut A;(\aut A_R)_{R\in\Rel}\bigr)$ of tree automata such that there exists a bijective \emph{naming function} $\mu\colon A\to L(\aut A)$ with the property that $\aut A_R$ recognises $\mu(R^{\struct A})$ for each $R\in\Rel$. A $\tau$-structure is \emph{tree-automatic} if it admits a tree-automatic presentation.
\end{definition}

\noindent In the situation above, the structure $\mu(\struct A)=\bigl(\mu(A);(\mu(R^{\struct A}))_{R\in\Rel}\bigr)$ is isomorphic to $\struct A$ and called a \emph{tree-automatic copy} of $\struct A$.

\begin{theorem}[Blumensath \cite{Blu99}]
Let $\struct A$ be a tree-automatic structure, $\bar{\aut A}$ a tree-automatic presentation of $\struct A$, $\mu$ the corresponding naming function, and $\phi(\bar x)$ an $\FOinf$-formula over~$\tau$. Then the relation $\mu(\phi^{\struct A})$ is automatic and one can compute a tree automaton recognising it from $\bar{\aut A}$ and $\phi$.
\end{theorem}

\begin{corollary}[Blumensath \cite{Blu99}]
Every tree-automatic structure possesses a decidable $\FOinf$-theory.
\end{corollary}

\section{Delhomm\'e's Decomposition Technique}

In this section, we present the decomposition technique Delhomm\'e used to show that every tree-automatic ordinal is below $\omega^{\omega^\omega}$.

\subsection{Sum and Box Augmentations and the Decomposition Theorem}

The central notions of Delhomm\'e's technique are sum augmentations and box augmentations.


\begin{definition}
\label{def:sum_augmentation}
A $\tau$-structure $\struct A$ is a \emph{sum augmentation} of $\tau$\nobreakdash-structures $\struct B_1,\dotsc,\allowbreak\struct B_n$ if there exists a finite partition $A=A_1\uplus\dotsb\uplus A_n$ of $\struct A$ such that $\struct A\restrict{A_i}\cong\struct B_i$ for each $\run i1n$.
\end{definition}

\begin{example}
\label{ex:sum_augmentation}
Let $\struct B_1,\dotsc,\struct B_n$ be linear orderings and $\struct A$ a linearisation of the partial ordering $\struct B_1\amalg\dotsb\amalg\struct B_n=\bigl(\biguplus_{1\leq i\leq n} B_i;\preceq)$ with $x\preceq y$ iff $x,y\in B_i$ and $x\leq^{\struct B_i} y$ for some $i$. Then $\struct A$ is a sum augmentation of $\struct B_1,\dotsc,\struct B_n$.
\end{example}

\begin{remark}
\label{rem:sum_augmentation}
Suppose a linear ordering $\struct A=(A;\leq^{\struct A})$ is a sum augmentation of $\struct B_1,\dotsc,\allowbreak\struct B_n$. First, each $\struct B_i$ can be embedded into $\struct A$ and hence is a linear ordering itself. Moreover, if $\struct A$ is a well-ordering, then each $\struct B_i$ is a well-ordering too. Second, $\struct A$ is isomorphic to a linearisation of $\struct B_1\amalg\dotsb\amalg\struct B_n$.
\end{remark}

\begin{definition}
\label{def:box_augmentation}
A $\tau$-structure $\struct A$ is a \emph{box augmentation} of $\tau$-structures $\struct B_1,\dotsc,\allowbreak\struct B_n$ if there exists a bijection $f\colon B_1\times\dotsb\times B_n\to A$ such that for all $\run j1n$ and $\bar x\in \prod_{1\leq i\leq n,i\not=j} B_i$ the map
\begin{equation*}
	f_{j,\bar x}\colon B_j\to A,b\mapsto (x_1,\dotsc,x_{j-1},b,x_{j+1},\dotsc,x_n)
\end{equation*}
is an embedding of $\struct B_j$ into $\struct A$.
\end{definition}

\begin{example}
\label{ex:box_augmentation}
Let $\struct B_1,\dotsc,\struct B_n$ be linear orderings and $\struct A$ a linearisation of the partial ordering $\struct B_1\times\dotsb\times\struct B_n=\bigl(B_1\times\dotsb\times B_n;\preceq)$ with $\bar x\preceq \bar y$ iff $x_i \leq^{\struct B_i} y_i$ for all $\run i1n$. Then $\struct A$ is a box augmentation of $\struct B_1,\dotsc,\struct B_n$.
\end{example}

\begin{remark}
\label{rem:box_augmentation}
Suppose a linear ordering $\struct A$ is a box augmentation of $\struct B_1,\dotsc,\allowbreak\struct B_n$. First, each $\struct B_i$ can be embedded into $\struct A$ and hence is a linear ordering itself. Moreover, if $\struct A$ is a well-ordering, then each $\struct B_i$ is a well-ordering too. Second, the bijection $f$ from Definition~\ref{def:box_augmentation} above is an isomorphism between a linearisation of $\struct B_1\times\dotsb\times\struct B_n$ and~$\struct A$.
\end{remark}

\noindent Since the concept of box augmentations is too general for our purposes, we need to restrict it. In the  following definition, an \emph{$R$-colouring} of a $\tau$\nobreakdash-structure $\struct B$ is a map ${c\colon B^{\ar(R)}\to Q}$ into a finite set $Q$ such that $c(\bar t)\in c(R^{\struct B})$ iff $\bar t\in R^{\struct B}$ for all $\bar t\in B^{\ar(R)}$.

\begin{definition}
\label{def:tame_box_augmentation}
The box augmentation in Definition~\ref{def:box_augmentation} is a \emph{tame box augmentation} if for each $R\in\Rel$ the following condition holds: For every $\run i1n$ there exists an $R$-colouring $c_i\colon B_i^{\ar(R)}\to Q_i$ of $\struct B_i$ such that the map
\begin{equation*}
	\prod\nolimits_{1\leq i\leq n} Q_i,
	\bigl(f(\bar x_1),\dotsc,f(\bar x_r)\bigr)\mapsto
	\bigl(c_i(x_{1,i},\dotsc,x_{r,i})\bigr){}_{\run i1n}
\end{equation*}
is an $R$-colouring of $\struct A$.
\end{definition}

\begin{remark}
Suppose a linear ordering $\struct A$ is a tame box augmentation of $\struct B_1,\dotsc,\allowbreak\struct B_n$. For each $\run i1n$ let $c_i\colon B_i^2\to Q_i$ be the corresponding $\leq$-colouring of $\struct B_i$. Without loss of generality, assume that the $Q_i$ all are the same set, say $\{1,\dotsc,m\}$. For each $\run i1n$ consider the structure $\struct C_i=\bigl(B_i;R_1^{\struct C_i},\dotsc,R_m^{\struct C_i}\bigr)$ with $R_j^{\struct C_i}=c_i^{-1}(j)$. Then the $R_j^{\struct C_j}$ form a finite partition of $B_i^2$ which is compatible with $\leq^{\struct B_i}$. Finally, the ordering $\struct A$ is a generalised product---in the sense of Feferman and Vaught---of the structures $\struct C_1,\dotsc,\struct C_n$ where only atomic formulae are used.
\end{remark}

\noindent More generally, the very essence of the notion of a tame box augmentation is to first partition all relations as well as their complements and to take a generalised product afterwards.

\begin{remark}
\label{rem:tame_box_augmentation}
If $\struct A$ is a tame box augmentation of $\struct B_1,\dotsc,\struct B_n$ and ${X_i\subseteq B_i}$ for each~$i$, then $\struct A\restrict{f(X_1\times\dotsb\times X_n)}$ is tame box augmentation of $\struct B_1\restrict{X_1},\dotsc,\struct B_n\restrict{X_n}$ via the bijection ${f\restrict{(X_1\times\dotsb\times X_n)}}$.
\end{remark}

\noindent In the situations of Definitions~\ref{def:sum_augmentation}, \ref{def:box_augmentation}, and~\ref{def:tame_box_augmentation} we also say that the structures $\struct B_1,\dotsc,\struct B_n$ form a \emph{sum decomposition} respectively a \emph{(tame) box decomposition} of $\struct A$. The decomposition theorem for tree-automatic structures is the following, whose proof is postponed to Section~\ref{sec:proof_thm:delhomme}.

\begin{theorem}[Delhomm\'e \cite{Del04}]
\label{thm:delhomme}
Let $\struct A$ be a tree-automatic $\tau$-structure and $\phi(x,y_1,\dotsc,y_n)$ an $\FOinf$-formula over $\tau$. Then there exists a finite set $\class S_\phi^{\struct A}$ of tree-automatic $\tau$\nobreakdash-structures such that for all $\bar s\in A^n$ the structure $\struct A\restrict{\phi^{\struct A}(\cdot,\bar s)}$ is a sum augmentation of tame box augmentations of elements from $\class S_\phi^{\struct A}$.
\end{theorem}

\noindent For now, suppose that $\class C$ is a class of $\tau$-structures ranked by $\nu$, i.e., $\nu$ assigns to each structure $\struct A\in\class C$ an ordinal $\nu(\struct A)$, its \emph{$\nu$-rank}, which is invariant under isomorphism. An ordinal $\alpha$ is \emph{$\nu$-sum-indecomposable} if for any structure $\struct A\in\class C$ with $\nu(\struct A)=\alpha$ every sum decomposition $\struct B_1,\dotsc,\struct B_n$ of $\struct A$ contains a component $\struct B_i$ with $\struct B_i\in\class C$ and $\nu(\struct B_i)=\alpha$. Similarly, we define \emph{$\nu$-(tame-)box-indecomposable} ordinals. Notice that every $\nu$-box-indecomposable ordinal is also $\nu$-tame-box-indecomposable. The following corollary is a direct consequence of Theorem~\ref{thm:delhomme}.

\begin{corollary}[Delhomm\'e \cite{Del04}]
\label{cor:delhomme_indecomposability}
Let $\class C$ be a class of $\tau$-structures ranked by $\nu$, $\struct A$ a tree-automatic $\tau$-structure, and $\phi(x,y_1,\dotsc,y_n)$ an $\FOinf$-formula over $\tau$. Then there are only finitely many ordinals $\alpha$ which are simultaneously $\nu$-sum-indecomposable as well as $\nu$\nobreakdash-tame-box-indecomposable and admit a $\bar s\in A^n$ with $\struct A\restrict{\phi^{\struct A}(\cdot,\bar s)}\in\class C$ and $\nu\bigl(\struct A\restrict{\phi^{\struct A}(\cdot,\bar s)}\bigr)=\alpha$.
\end{corollary}

\begin{proof}
Let $\class S_\phi^{\struct A}$ be the finite set of structures which exists by Theorem~\ref{thm:delhomme}. Consider an ordinal $\alpha$ which is $\nu$-sum-indecomposable as well as $\nu$-tame-box-indecomposable and admits a tuple $\bar s\in A^n$ with $\struct A\restrict{\phi^{\struct A}(\cdot,\bar s)}\in\class C$ and $\nu\bigl(\struct A\restrict{\phi^{\struct A}(\cdot,\bar s)}\bigr)=\alpha$. Then there exists a tame box decomposition $\struct B_1,\dotsc,\struct B_m$ of $\struct A\restrict{\phi^{\struct A}(\cdot,\bar s)}$ such that each $\struct B_i$ is a sum augmentation of elements from $\class S_\phi^{\struct A}$. Since $\alpha$ is $\nu$-tame-box-indecomposable, there is an $\elem{i_0}1m$ such that $\struct B_{i_0}\in\class C$ and $\nu(\struct B_{i_0})=\alpha$. Moreover, there exists a sum decomposition $\struct C_1,\dotsc,\struct C_n$ of $\struct B_{i_0}$ such that $\struct C_j\in\class S_\phi^{\struct A}$ for each $\run j1n$. As $\alpha$ is also $\nu$-sum-indecomposable, there is a $\elem{j_0}1n$ such that $\struct C_{j_0}\in\class C$ and $\nu(\struct C_{j_0})=\alpha$.

In particular, $\class S_\phi^{\struct A}$ contains a structure $\struct B$ with $\struct B\in\class C$ and $\nu(\struct B)=\alpha$. Since $\class S_\phi^{\struct A}$ is finite, there are only finitely many ordinals $\alpha$ of the type under consideration.
\end{proof}

\subsection{Tree-Automatic Ordinals}

In order to prove that every tree-automatic ordinal is strictly below $\omega^{\omega^\omega}$, we apply Corollary~\ref{cor:delhomme_indecomposability} to the class of all well-orderings and rank each well-ordering $\struct A$ by its order type $\tp(\struct A)$. To identify the $\tp$-sum-indecomposable and $\tp$-box-indecomposable ordinals, we need the natural sum and product.

Due to the Cantor normal form, every ordinal can be regarded as a polynomial in $\omega$ with natural numbers as coefficients and ordinals as exponents. Intuitively, the natural sum of two ordinals is formed by adding the corresponding polynomials and the natural product by multiplying the polynomials whereby exponents are added using the natural sum. Formally, let $\alpha=\sum_{i=1}^{i=n} \omega^{\gamma_i}k_i$ and $\beta=\sum_{i=1}^{i=n} \omega^{\gamma_i}\ell_i$ with $\gamma_1>\dotsb>\gamma_n\geq 0$ and $k_1,\dotsc,k_n,\ell_1,\dotsc,\ell_n\in\Nat$ be two ordinals in Cantor normal form. The \emph{natural sum} $\alpha\oplus\beta$ and the \emph{natural product} $\alpha\otimes\beta$ are defined by
\begin{equation*}
	\alpha\oplus\beta = \sum\nolimits_{i=1}^{i=n} \omega^{\gamma_i}(k_i+\ell_i)
	\qquad\text{and}\qquad
	\alpha\otimes\beta =
	\bigoplus\nolimits_{i,j=1}^{i,j=n} \omega^{\gamma_i\oplus\gamma_j} k_i\ell_j\,.
\end{equation*}
Compared with the usual addition and multiplication of ordinals, both operations are commutative and strictly monotonic in both arguments and $\otimes$ distributes over $\oplus$. The following theorem is an adaption of results in \cite{Car42} to our setting.

\begin{theorem}[Caruth \cite{Car42}]
\label{thm:caruth}
Let $\alpha$ and $\beta_1,\dotsc,\beta_n$ be ordinals.
\begin{enumerate}
\item If $\alpha$ is a sum augmentation of $\beta_1,\dotsc,\beta_n$, then $\alpha \leq \beta_1\oplus\dotsb\oplus\beta_n$.
\item If $\alpha$ is a box augmentation of $\beta_1,\dotsc,\beta_n$, then $\alpha \leq \beta_1\otimes\dotsb\otimes\beta_n$.
\end{enumerate}
\end{theorem}

\begin{corollary}
\label{cor:indecomposable_ordinals}
Let $\alpha$ be an ordinal. Then $\omega^\alpha$ is $\tp$-sum-indecomposable and $\omega^{\omega^\alpha}$ is $\tp$-box-indecomposable.
\end{corollary}

\begin{proof}
Let $\beta_1,\dotsc,\beta_n$ be a sum decomposition of $\omega^\alpha$. Then $\beta_i\leq\omega^\alpha$ for each $i$. If $\beta_i<\omega^\alpha$ for all $i$, then $\beta_1\oplus\dotsb\oplus\beta_n<\omega^\alpha$. This contradicts Theorem~\ref{thm:caruth}~(1).

Now, let $\beta_1,\dotsc,\beta_n$ be a box decomposition of $\omega^{\omega^\alpha}$. Then $\beta_i\leq\omega^{\omega^\alpha}$ for each $i$. By contradiction, assume $\beta_i<\omega^{\omega^\alpha}$ for all $i$. Since $\omega^{\omega^\alpha}$ is a limit ordinal, there are $\gamma_i<\omega^\alpha$ with $\beta_i<\omega^{\gamma_i}$ and hence
\begin{equation*}
	\beta_1\otimes\dotsb\otimes\beta_n <
	\omega^{\gamma_1\oplus\dotsb\oplus\gamma_n} <
	\omega^{\omega^\alpha}\,.
\end{equation*}
This contradicts Theorem~\ref{thm:caruth}~(2).
\end{proof}

\noindent Finally, Corollaries~\ref{cor:delhomme_indecomposability} and~\ref{cor:indecomposable_ordinals} imply that any tree-automatic ordinal is strictly less than $\omega^{\omega^\omega}$. The main ingredient for the converse implication is the following lemma.

\begin{lemma}
\label{lemma:ordinals_are_tree_automatic}
For each $k\in\Nat$ the ordinal $\omega^{\omega^k}$ admits a tree-automatic presentation over a unary alphabet $\Sigma$.
\end{lemma}

\begin{proof}
We proceed by induction on $k$.

\subparagraph*{Base case.} $k=0$.\\
The map $\mu\colon\omega\to\trees\Sigma$ which assigns to $n\in\omega$ the unique tree $\mu(n)$ with ${\dom\bigl(\mu(n)\bigr) = \{0\}^{<n}}$ can be used as naming function for a tree-automatic presentation of $\omega$.

\subparagraph*{Inductive step.} $k>0$.\\
We regard $\omega^{\omega^k}$ as the length-lexicographically ordered set of all maps $f\colon\omega\to\omega^{\omega^{k-1}}$ which are zero almost everywhere. Let $\nu$ be the naming function corresponding to the tree-automatic presentation of $\omega^{\omega^{k-1}}$ which exists by induction. We define a map $\mu\colon\omega^{\omega^k}\to\trees\Sigma$ by letting $\mu(f)$ be the unique tree with
\begin{equation*}
	\dom\bigl(\mu(f)\bigr) =
		\bigcup_{0\leq i<n} \bigl\{0^i\bigr\} \cup \bigl\{0^i1\}\dom\bigl(\nu\bigl(f(i)\bigr)\bigr)\,,
\end{equation*}
where $n\in\omega$ is minimal with $f(m)=0$ for all $m\geq n$. This map can be used as naming function for a tree-automatic presentation of $\omega^{\omega^k}$.
\end{proof}

\begin{corollary}[Delhomm\'e \cite{Del04}]
\label{cor:main_ordinal}
An ordinal $\alpha$ is tree-automatic if, and only if,
\begin{equation*}
	\alpha < \omega^{\omega^\omega}\,.
\end{equation*}
\end{corollary}

\begin{proof}
By contradiction, assume there exists a tree-automatic ordinal ${\alpha\geq\omega^{\omega^\omega}}$. Consider $\phi(x,y)=x\leq y\land x\not=y$. Clearly, $\phi^\alpha\bigl(\cdot,\beta)=\beta$ for every $\beta\in\alpha$. In particular, $\tp\bigl(\alpha\restrict{\phi^\alpha\bigl(\cdot,\omega^{\omega^d}\bigr)}\bigr)=\omega^{\omega^d}$ for each $d\in\Nat$. Since these ordinals $\omega^{\omega^d}$ are $\tp$-sum-indecomposable as well as $\tp$-box-indecomposable, this contradicts Corollary~\ref{cor:delhomme_indecomposability}.

Now, let $\alpha<\omega^{\omega^\omega}$ be some ordinal. There exists a $k\in\Nat$ such that $\alpha<\omega^{\omega^k}$. By Lemma~\ref{lemma:ordinals_are_tree_automatic}, $\omega^{\omega^k}$ is tree-automatic. Finally, $\alpha$ is $\FO$-definable with one parameter in $\omega^{\omega^k}$ and hence tree-automatic.
\end{proof}

\subsection{Proof of the Decomposition Theorem}
\label{sec:proof_thm:delhomme}

We conclude this section by providing a proof of Theorem~\ref{thm:delhomme}.

\begin{proof}[Proof of Theorem~\ref{thm:delhomme}]
\newcommand{\state}[1]{\llbracket #1\rrbracket}
Let $\bigl(\aut A;(\aut A_R)_{R\in\Rel}\bigr)$ be a tree-automatic presentation of $\struct A$ with $L(\aut A)\subseteq\trees\Sigma$. To keep notation simple, we assume that the corresponding naming function $\mu\colon A\to L(\aut A)$ is the identity, i.e., $\struct A$ is identified with its tree-automatic copy $\mu(\struct A)$. For $R\in\Rel$ let $Q_R$ be the set of states of $\aut A_R$. Moreover, let $\aut A_\phi$ be a tree automaton recognising $\phi^{\struct A}$ and $Q_\phi$ its set of states. For each $t\in\trees\Sigma$ and all $r\geq 1$ we put $\otimes_r t=\otimes(t,\dotsc,t)\in\trees{\Sigma_\Box^r}$, where the convolution is made up of $r$ copies of $t$. We further define a tree $\boxtimes_n t=(t,\emptyset,\dotsc,\emptyset)\in\trees{\Sigma_\Box^{1+n}}$, where the number of empty trees $\emptyset$ in the convolution is $n$. To simplify notation even more, we put
\begin{equation*}
	\state{t}_\phi = \aut A_\phi(\boxtimes_n t)
	\qquad\text{and}\qquad
	\state{t}_R = \aut A_R\bigl(\otimes_{\ar(R)} t\bigr)
\end{equation*}
for every $t\in\trees\Sigma$ and $R\in\Rel$.

Consider the set
\begin{equation*}
	\Gamma = \prod_{R\in\{\phi\}\uplus\Rel} Q_R \times \prod_{R\in\Rel} 2^{Q_R}\,.
\end{equation*}
For each $\gamma=\bigl((q_R)_{R\in\{\phi\}\uplus\Rel},(P_R)_{R\in\Rel}\bigr)\in\Gamma$ we define a structure $\struct S_\gamma$ by
\begin{equation*}
	\univ{\struct S_\gamma} = S_\gamma = \Set{ t\in\trees\Sigma |
	\text{$\state{t}_\phi=q_\phi$ and $\state{t}_R=q_R$ for each $R\in\Rel$} }
\end{equation*}
and
\begin{equation*}
	R^{\struct S_\gamma} = \Set{ \bar t\in S_\gamma^{\ar(R)} |
		\aut A_R(\otimes\bar t)\in P_R }\ \text{for $R\in\Rel$.}
\end{equation*}
Clearly, $\struct S_\gamma$ is a tree-automatic copy of itself. Finally, we put
\begin{equation*}
	\class S_\phi^{\struct A} = \Set{ \struct S_\gamma | \gamma\in\Gamma }\,.
\end{equation*}
Obviously, this set is finite.

For the rest of this proof, we fix some parameters $\bar s=(s_1,\dotsc,s_n)\in A^n$ and put ${D=\bigcup_{1\leq i\leq n} \dom(s_i)}$. The \emph{$\bar s$-type} of a tree $t\in\trees\Sigma$ is the tuple
\begin{equation*}
	\tp_{\bar s}(t)=
	\bigl(t\restrict D,U,(\rho_R)_{R\in\{\phi\}\uplus\Rel}\bigr)\,,
\end{equation*}
where $t\restrict D\in\trees\Sigma$ is the restriction of $t$ to the tree domain $\dom(t)\cap D$, $U=\dom(t)\cap\partial D$, and $\rho_R\colon U\to Q_R,u\mapsto\state{t\restrict u}_R$ for each $R\in\{\phi\}\uplus\Rel$. Observe that
\begin{equation*}
	\otimes(t,\bar s) = \otimes(t\restrict D,\bar s)\bigl[(u/\boxtimes_n t\restrict u)_{u\in U}\bigr]
\end{equation*}
and hence
\begin{equation}
	\label{eq:tp_saturates_phi}
	\aut A_\phi\bigl(\otimes(t,\bar s)\bigr) = \aut A_\phi\bigl(\otimes(t\restrict D,\bar s),\rho_\phi\bigr)\,,
\end{equation}
i.e., whether $t\in\phi^\struct A(\cdot,\bar s)$ is valid can be determined from $\tp_{\bar s}(t)$. Since $D$ is finite, there are only finitely many distinct $\bar s$-types. Consequently, the equivalence relation $\sim_{\bar s}$ on $T_\Sigma$ defined by $t\sim_{\bar s} t'$ iff $\tp_{\bar s}(t)=\tp_{\bar s}(t')$ has finite index. Due to Eq.~\eqref{eq:tp_saturates_phi}, $\phi^{\struct A}(\cdot,\bar s)$ is a union of $\sim_{\bar s}$-classes. Say $B_1,\dotsc,B_m\subseteq\phi^{\struct A}(\cdot,\bar s)$ are these $\sim_{\bar s}$-classes, then $\struct A\restrict{\phi^{\struct A}(\cdot,\bar s)}$ is a sum augmentation of $\struct A\restrict{B_1},\dotsc,\struct A\restrict{B_m}$. Thus, it remains to show that $\struct A\restrict B$ is a tame box augmentation of elements from $\class S_\phi^{\struct A}$ for each $\sim_{\bar s}$-class $B\subseteq\phi^{\struct A}(\cdot,\bar s)$.

Therefore, fix some $\sim_{\bar s}$-class $B\subseteq\phi^{\struct A}(\cdot,\bar s)$, let $\vartheta=\bigl(t_D,U,(\rho_R)_{R\in\{\phi\}\uplus\Rel}\bigr)$ be the corresponding $\bar s$-type, and put $\struct B=\struct A\restrict B$. For $u\in U$ we define

\begin{equation*}
	\gamma(\vartheta,u) = 
	\bigl((\rho_R(u))_{R\in\{\phi\}\uplus\Rel},(P_R(u))_{R\in\Rel}\bigr)
	\in\Gamma
\end{equation*}
by
\begin{equation*}
	P_R(u) = \Set{ q\in Q_R |
	\aut A_R\bigl(\otimes_{\ar(R)} t_D,\rho_R[u\mapsto q]\bigr)\in F_R }\ \text{for $R\in\Rel$,}
\end{equation*}
where $F_R\subseteq Q_R$ is the set of accepting states of $\aut A_R$.  Let $u_1,\dotsc,u_m$ be an enumeration of the elements of $U$ and put $\struct C_i=\struct S_{\gamma(\vartheta,u_i)}$ for $\run i1m$. Next, we show that $\struct B$ is a tame box augmentation of $\struct C_1,\dotsc,\struct C_m$.

First, observe that
\begin{equation*}
	f\colon C_1\times\dotsb\times C_m\to\trees\Sigma,
	(x_1,\dotsc,x_m)\mapsto t_D[u_1/x_1,\dotsc,u_m/x_m]
\end{equation*}
is injective. Some $t\in\trees\Sigma$ is contained in the image of $f$ if, and only if, $t\restrict D=t_D$, ${\dom(t)\cap\partial D=U}$, and $t\restrict{u_i}\in C_i$ for each $\run i1m$. The latter is equivalent to $\tp_{\bar s}(t)=\vartheta$ and hence $f$ is a bijection $f\colon C_1\times\dotsb\times C_m\to B$. Fix some $\run j1m$ and $\bar x\in\prod_{1\leq i\leq m,i\not=j} C_i$ and let
\begin{equation*}
	f_{j,\bar x}\colon C_j\to B,
	t\mapsto f(x_1,\dotsc,x_{j-1},t,x_{j+1},\dotsc,x_m)\,.
\end{equation*}
Consider $R\in\Rel$ and $r=\ar(R)$. For all $\bar t\in C_j^r$ we have
\begin{equation*}
	\otimes f_{j,\bar x}(\bar t) =
	(\otimes_r t_D)\bigl[(u_i/\otimes_r x_i)_{1\leq i\leq m,i\not=j},
	u_j/\otimes\bar t\bigr]
\end{equation*}
and hence
\begin{equation*}
	\aut A_R\bigl(\otimes f_{j,\bar x}(\bar t)\bigr) =
	\aut A_R\bigl(\otimes_r t_D,
		\rho_R\bigl[u_j\mapsto \aut A_R(\otimes\bar t)\bigr]\bigr)\,.
\end{equation*}
This leads to the following chain of equivalences
\begin{align*}
	f_{j,\bar x}(\bar t)\in R^{\struct B}
	&\quad\Longleftrightarrow\quad
		\aut A_R\bigl(\otimes f_{j,\bar x}(\bar t)\bigr)\in F_R \\
	&\quad\Longleftrightarrow\quad
		\aut A_R\bigl(\otimes_r t_D,
			\rho_R\bigl[u_j\mapsto \aut A_R(\otimes\bar t)\bigr]\bigr)\in F_R \\
	&\quad\Longleftrightarrow\quad
		\aut A_R(\otimes\bar t)\in P_R(u_j) \\
	&\quad\Longleftrightarrow\quad
		\bar t\in R^{\struct C_j}\,,
\end{align*}
which shows that $\struct B$ is a box augmentation of $\struct C_1,\dotsc,\struct C_m$. It remains to show that this box augmentation is tame.

Therefore, fix some $R\in\Rel$, put $r=\ar(R)$, and notice that the map
\begin{equation*}
	c_i\colon C_i^r\to Q_R,\bar t\mapsto \aut A_R(\otimes\bar t)
\end{equation*}
is an $R$-colouring of $\struct C_i$ for each $\run i1m$. We have to show that
\begin{equation*}
	c\colon B^r\to Q_R^m,
	\bigl(f(\bar x_1),\dotsc,f(\bar x_r)\bigr)\mapsto
	\bigl(c_i(x_{1,i},\dotsc,x_{r,i})\bigr){}_{1\leq i\leq m}
\end{equation*}
is an $R$-colouring of $\struct B$. Consider the map
\begin{equation*}
	h\colon Q_R^m\to Q_m,
	(q_1,\dotsc,q_m)\mapsto \aut A_R\bigl(\otimes_r t_D,
		\set{ u_i\mapsto q_i | 1\leq i\leq m }\bigr)\,.
\end{equation*}
For every  $\bar t\in B^r$ we obtain $h\bigl(c(\bar t)\bigr)=\aut A_R(\otimes\bar t)$ and hence $h\circ c$ is an $R$-colouring of $\struct B$. Consequently, $c$ is an $R$-colouring of $\struct B$ as well.
\end{proof}

\section{Tree-Automatic Linear Orderings}

The objective of this section is to prove our main result, namely Theorem~\ref{thm:main}, which states that every tree-automatic linear ordering has $\rankFC$\nobreakdash-rank below $\omega^\omega$. Due to the fact that every countable linear ordering is a dense sum of scattered linear orderings, the proof is essentially an application of Corollary~\ref{cor:delhomme_indecomposability} to the class of countable scattered linear orderings ranked by $\rankVD_*$, a variation of the $\rankFC$-rank. Since it is already known that every ordinal is $\rankVD_*$-sum-indecomposable \cite{KRS05}, the major part of this section is devoted to identifying the $\rankVD_*$-tame-box-indecomposable ordinals.

\subsection{Linear Orderings and the $\rankFC$-rank}

A \emph{(linear) ordering} is a structure $\struct A=(A;\leq^{\struct A})$ where $\leq^{\struct A}$ is a \emph{non-strict} linear order on~$A$. Sometimes we use the corresponding \emph{strict} linear order $<^{\struct A}$. If $\struct A$ is clear from the context we omit the superscript $\struct A$. An \emph{interval} in $\struct A$ is a subset $I\subseteq A$ such that $x<z<y$ implies $z\in I$ for all $x,y\in I$ and $z\in A$. For $x,y\in A$ the \emph{closed interval} $[x,y]_{\struct A}$ in $\struct A$ is the set $\set{ z\in A | x\leq z\leq y}$ if $x\leq y$ and the set $\set{ z\in A | y\leq z\leq x}$ if $x>y$.

\begin{definition}
A \emph{condensation (relation)} on a linear ordering $\struct A$ is an equivalence relation $\sim$ on $A$ such that each $\sim$-class is an interval of $\struct A$.
\end{definition}

\noindent For two subsets $X,Y\subseteq A$ we write $X\ll Y$ if $x<y$ for all $x\in X$ and $y\in Y$. If $\sim$ is a condensation on $\struct A$, the set $A/\mathord{\sim}$ of all $\sim$-classes is (strictly) linearly ordered by $\ll$. We denote the corresponding linear ordering by $\struct A/\mathord{\sim}$. An example of a condensation is the relation $\sim$ with $x\sim y$ iff the closed interval $[x,y]_{\struct A}$ in $\struct A$ is finite. The ordering $\struct A/\mathord{\sim}$ is obtained from $\struct A$ by identifying points which are only finitely far away from each other. If this process is transfinitely iterated, it eventually becomes stationary. Intuitively, the $\rankFC$-rank of $\struct A$ is the ordinal $\alpha$ counting the number of steps which are necessary to reach this fix point.

\begin{definition}
Let $\struct A$ be a linear ordering. For each ordinal $\alpha$ a condensation $\sim_\alpha^{\struct A}$ on $\struct A$ is defined by transfinite induction:
\begin{enumerate}
\item $\sim_0^{\struct A}$ is the identity relation on $\struct A$,
\item for successor ordinals $\alpha=\beta+1$ let $x\sim_\alpha^{\struct A} y$ iff the interval $[\tilde x,\tilde y]_{\struct A/\mathord{\sim_\beta^{\struct A}}}$ in $\struct A/\mathord{\sim_\beta^{\struct A}}$ is finite, where $\tilde x$ and $\tilde y$ are the $\sim_\beta^{\struct A}$-classes of $x$ and $y$, and
\item for limit ordinals $\alpha$ let $x\sim_\alpha^{\struct A} y$ iff $x\sim_\beta^{\struct A} y$ for some $\beta<\alpha$.
\end{enumerate}
\end{definition}

\noindent For each ordering $\struct A$ there exists an ordinal $\alpha$ such that $\sim_\alpha^{\struct A}$ and $\sim_\beta^{\struct A}$ coincide for each $\beta\geq\alpha$. More precisely, every ordinal $\alpha$ whose cardinality is greater than the one of $\struct A$ has this property. Theorem 5.9 in \cite{Ros82} ascertains that if $\struct A$ is countable then $\alpha$ can be chosen countable as well.

\begin{definition}
The \emph{$\rankFC$-rank} of a linear ordering $\struct A$, denoted by $\rankFC(\struct A)$, is the least ordinal $\alpha$ such that $\sim_\alpha^{\struct A}$ and $\sim_\beta^{\struct B}$ coincide for each $\beta\geq\alpha$.
\end{definition}

\noindent For a linear ordering $\struct A$ and a subset $B\subseteq A$ we simply write $\rankFC(B)$ for $\rankFC(\struct A\restrict B)$. The following theorem is the main result of this article.

\begin{theorem}
\label{thm:main}
Let $\struct A$ be a tree-automatic linear ordering. Then
\begin{equation*}
	\rankFC(\struct A)< \omega^\omega\,.
\end{equation*}
\end{theorem}

\noindent Since $\rankFC(\alpha)\leq\beta$ if, and only if, $\alpha\leq\omega^\beta$ for all countable ordinals $\alpha$ and $\beta$, Theorem~\ref{thm:main} above yields another proof of the fact that every tree-automatic ordinal is strictly less than $\omega^{\omega^\omega}$ (cf. Corollary~\ref{cor:main_ordinal}).

\subsection{Scattered Linear Orderings and the $\rankVD$-rank}

\emph{Throughout the rest of this paper, we consider only countable linear orderings.} A linear ordering $\struct A$ is \emph{scattered} if the ordering $(\mathbb{Q};<)$ of the rationals cannot be embedded into $\struct A$, or equivalently, if there exists an ordinal $\alpha$ such that $\struct A/\mathord{\sim_\alpha^{\struct A}}$ contains exactly one element (cf. Chapter 5 in \cite{Ros82}). Examples of scattered orderings include the natural numbers $\omega=(\Nat;\leq)$, the reversed natural numbers $\omega^*=(\Nat;\geq)$, the integers $\zeta=(\mathbb{Z};\leq)$, and the finite linear orderings $\mathbf{n}=\bigl(\{1,\dotsc,n\};\leq\bigr)$ for $n\in\Nat$. Furthermore, every ordinal is scattered.

For an ordering $\struct I$ the \emph{$\struct I$-sum} of an $I$-indexed family $(\struct A_i)_{i\in I}$ of orderings is the linear ordering
\begin{equation*}
	\struct A = \sum\nolimits_{i\in\struct I} \struct A_i
\end{equation*}
defined by $A = \biguplus_{i\in I} A_i$ and $x\leq^{\struct A} y$ iff $x,y\in A_i$ and $x\leq^{\struct A_i} y$ for some $i\in I$ or $x\in A_i$ and $y\in A_j$ for some $i,j\in I$ with $i<^{\struct I} j$.
If $\struct I$ is finite, say $\struct I=\mathbf{n}$, we write $\struct A_1+\dotsb+\struct A_n$ for $\sum_{i\in\mathbf{n}} \struct A_i$.

Next, we introduce the class of very discrete linear orderings and their connection to the scattered linear orderings.

\begin{definition}
For each countable ordinal $\alpha$ the class $\VD_\alpha$ of linear orderings is defined by transfinite induction:
\begin{enumerate}
\item $\VD_0 = \{\mathbf{0},\mathbf{1}\}$, and
\item for $\alpha>0$ the class $\VD_\alpha$ contains all finite sums, $\omega$-sums, $\omega^*$-sums, and $\zeta$-sums of elements from $\VD_{<\alpha}=\bigcup_{\beta<\alpha}\VD_\alpha$.
\end{enumerate}
The class $\VD$ of \emph{very discrete} linear orderings is the union of all classes $\VD_\alpha$. The \emph{$\rankVD$-rank} of some $\struct A\in\VD$, denoted by $\rankVD(\struct A)$, is the least ordinal $\alpha$ with $\struct A\in\VD_\alpha$.
\end{definition}

\noindent The following result is due to Hausdorff and Theorem~5.24 in \cite{Ros82}.

\begin{theorem}[Hausdorff \cite{Hau08}]
A countable linear ordering $\struct A$ is scattered if, and only if, it is contained in $\VD$. In case $\struct A$ is scattered,
\begin{equation*}
	\rankFC(\struct A) = \rankVD(\struct A)\,.
\end{equation*}
\end{theorem}

\noindent In order to formulate the intermediate steps of our proof of Theorem~\ref{thm:main}, we need a slight variation of the $\rankVD$-rank \cite{KRS05}.

\begin{definition}
The \emph{$\rankVD_*$-rank} of a scattered linear ordering $\struct A$, denoted by $\rankVD_*(\struct A)$, is the least ordinal $\alpha$ such that $\struct A$ is a finite sum of elements from $\VD_\alpha$.
\end{definition}

\noindent The $\rankVD$-rank and the $\rankVD_*$-rank of a scattered linear ordering $\struct A$ are closely related by the following inequality
\begin{equation}
	\label{eq:VD_inequality}
	\rankVD_*(\struct A)\leq \rankVD(\struct A)\leq\rankVD_*(\struct A)+1\,.
\end{equation}
The following lemma is very useful when reasoning about the ranks of scattered linear orderings. The first inequality is Lemma~5.14 in \cite{Ros82} and the second inequality is a trivial consequence of the first one.

\begin{lemma}
\label{lemma:VD_subordering}
Let $\struct A$ be a scattered linear ordering and $B\subseteq A$. Then
\begin{equation*}
	\rankVD(\struct A\restrict B) \leq \rankVD(\struct A)
	\qquad\text{and}\qquad
	\rankVD_*(\struct A\restrict B) \leq \rankVD_*(\struct A)\,.
\end{equation*}
\end{lemma}



\subsection{Sum and Box Augmentations of Scattered Linear Orderings}

Every sum decomposition of a scattered linear ordering $\struct A$ entirely consists of scattered linear orderings (cf. Remark~\ref{rem:sum_augmentation}). The relationship between the $\rankVD_*$-ranks of $\struct A$ and the components was established in \cite{KRS05}.

\begin{proposition}[Khoussainov, Rubin, Stephan \cite{KRS05}]
\label{prop:VD_sum}
Let $\struct A$ be a scattered linear ordering and a sum augmentation of $\struct B_1,\dotsc,\struct B_n$. Then
\begin{equation*}
	\rankVD_*(\struct A) =
	\max \bigl\{ \rankVD_*(\struct B_1),\dotsc,\rankVD_*(\struct B_n) \bigr\}\,.
\end{equation*}
\end{proposition}

\begin{corollary}
\label{cor:VD_sum_indecomposable}
Every countable ordinal is $\rankVD_*$-sum-indecomposable.
\end{corollary}

\noindent As already mentioned, we are mainly interested in the $\rankVD_*$-tame-box-indecomposable ordinals. The main tool for identifying them is Proposition~\ref{prop:VD_box} below whose proof is postponed to page~\pageref{proof:prop:VD_box}. Notice that Remark~\ref{rem:sum_augmentation} implies that  $\struct B_1,\dotsc,\struct B_n$ therein are scattered linear orderings.

\begin{proposition}
\label{prop:VD_box}
Let $\struct A$ be a scattered linear ordering and a tame box augmentation of $\struct B_1,\dotsc,\struct B_n$. Then
\begin{equation*}
	\rankVD_*(\struct A) \leq
	\rankVD_*(\struct B_1)\oplus\dotsb\oplus\rankVD_*(\struct B_n)\,.
\end{equation*}
\end{proposition}

\begin{corollary}
\label{cor:VD_box_indecomposable}
Every countable ordinal of the shape $\omega^\alpha$ is $\rankVD_*$-tame-box-indecomposable.
\end{corollary}

\begin{proof}
Let $\struct A$ be a scattered linear ordering with $\rankVD_*(\struct A)=\omega^\alpha$ and $\struct B_1,\dotsc,\struct B_n$ a tame box decomposition of $\struct A$. Since each $\struct B_i$ can be embedded into $\struct A$, Lemma~\ref{lemma:VD_subordering} yields ${\rankVD_*(\struct B_i)\leq\omega^\alpha}$. If $\rankVD_*(\struct B_i)<\omega^\alpha$ for each $i$, then
\begin{equation*}
	\rankVD_*(\struct B_1)\oplus\dotsb\oplus\rankVD_*(\struct B_n) <
	\omega^\alpha\,.
\end{equation*}
This contradicts Proposition~\ref{prop:VD_box}.
\end{proof}

\noindent As a first step towards the proof of Proposition~\ref{prop:VD_box} we provide two rather technical lemmas.

\begin{lemma}
\label{lemma:ramsey_sequence}
Let $\struct A$ be a linear ordering without a greatest element and $c\colon A^2\to Q$ a $\leq$-colouring of $\struct A$. Then there exist a strictly increasing, unbounded sequence $(a_i)_{i\in\Nat}$ in $\struct A$ and a colour $q\in Q$ such that $c(a_i,a_j)=q$ for all $i,j\in\Nat$ with $i<j$.
\end{lemma}

\begin{proof}
Since $\struct A$ has no greatest element, there exists a strictly increasing and unbounded sequence $(x_i)_{i\in\Nat}$ in $\struct A$. By Ramsey's theorem for infinite, undirected, edge coloured graphs there exist an infinite set $H\subseteq\Nat$ and a colour $q\in Q$ such that $c(x_i,x_j)=q_1$ for all $i,j\in H$ with $i<j$. Let $k_0<k_1<\dotsb$ be the increasing enumeration of all elements in $H$ and put $a_i=x_{k_i}$ for all $i\in\Nat$.
\end{proof}

\noindent Notice that the dual of this lemma holds as well and makes a statement about linear orderings without a least element and strictly decreasing, unbounded sequences. In the following lemma, the interval $(-\infty,a_0]_{\struct A}$ denotes the set of all $a\in A$ with $a\leq a_0$.

\begin{lemma}
\label{lemma:sequence_split}
Let $\struct A$ be an $\omega$-sum of elements from $\VD_{<\alpha}$ and $(a_i)_{i\in\Nat}$ a increasing sequence in $\struct A$. Then
\begin{equation*}
	\rankVD_*\bigl((-\infty,a_0]_{\struct A}\bigr)<\alpha
	\quad\text{and}\quad
	\rankVD_*\bigl((a_{k-1},a_k]_{\struct A}\bigr)<\alpha\ \text{for all $k\geq 1$.}
\end{equation*}
\end{lemma}

\begin{proof}
Let $\struct A=\sum_{i\in\omega} \struct A_i$ with $\struct A_i\in\VD_{<\alpha}$ for all $i\in\omega$. For each $k\in\omega$ there exists a unique $\ell\in\omega$ with $a_k\in A_\ell$. Then $(-\infty,a_k]_{\struct A}\subseteq A_0\cup\dotsb\cup A_\ell$ and hence
\begin{equation*}
	\rankVD_*\bigl((-\infty,a_k]_{\struct A}\bigr) \leq \rankVD_*(\struct A_0+\dotsb+\struct A_\ell)<\alpha\,.
\end{equation*}
Moreover, for $k\geq 1$ we have $\rankVD_*\bigl((a_{k-1},a_k]_{\struct A}\bigr)\leq\rankVD_*\bigl((-\infty,a_k]_{\struct A}\bigr)<\alpha$.
\end{proof}

\noindent Again, the dual of this statement which speaks about $\omega^*$-sums and decreasing sequences holds true. Basically, the proof of Proposition~\ref{prop:VD_box} proceeds by induction on $n$ and reduces thus to the case $n=2$. Proposition~\ref{prop:VD_box2} slightly rephrases the claim for $n=2$.

\begin{proposition}
\label{prop:VD_box2}
Let $\alpha$ and $\beta$ be ordinals, $\struct C$ a scattered linear ordering, and $\struct A$ and $\struct B$ form a tame box decomposition of $\struct C$ with $\rankVD_*(\struct A)\leq\alpha$ and $\rankVD_*(\struct B)\leq\beta$. Then
\begin{equation}
	\label{eq:VD_box2}
	\rankVD_*(\struct C) \leq \alpha\oplus\beta\,.
\end{equation}
\end{proposition}

\begin{proof}
We proceed by induction on $\alpha$ and $\beta$. To keep notation simple, we assume that the map $f\colon A\times B\to C$ from the definition of box augmentation is the identity, i.e., $C=A\times B$ and $\struct C$ is a linearisation of $\struct A\times\struct B$ (cf. Remark~\ref{rem:box_augmentation}).

Before delving into the induction, we perform a slight simplification. By definition, there exist $\struct A_1,\dotsc,\struct A_m\in\VD_\alpha$ and $\struct B_1,\dotsc,\struct B_n\in\VD_\beta$ such that $\struct A=\struct A_1+\dotsb+\struct A_m$ and $\struct B=\struct B_1+\dotsb+\struct B_n$. Since every $\zeta$-sum of linear orderings can be written as a sum of an $\omega$-sum and an $\omega^*$\nobreakdash-sum, we can assume that none of the $\struct A_i$ or $\struct B_j$ is constructed as a $\zeta$-sum. Obviously, $\struct C$ is a sum augmentation of the $m\cdot n$ orderings $\struct C\restrict{(A_i\times B_j)}$. By Proposition~\ref{prop:VD_sum}, it suffices to show 
\begin{equation*}
	\rankVD_*\bigl(\struct C\restrict{(A_i\times B_j)}\bigr)\leq
	\alpha\oplus\beta
\end{equation*}
for all $i$ and $j$. Since $\struct C\restrict{(A_i\times B_j)}$ is a tame box augmentation of $\struct A_i$ and $\struct B_j$, it remains to show Eq.~\eqref{eq:VD_box2} under the stronger assumptions that $\rankVD(\struct A)\leq\alpha$, $\rankVD(\struct B)\leq\beta$, and neither $\struct A$ nor $\struct B$ is constructed as a $\zeta$-sum.

\subparagraph*{Base case.} $\alpha=0$ or $\beta=0$.\\
If $\alpha=0$, then $\struct A\cong\mathbf{1}$ and $\struct C\cong\struct B$. Thus, $\rankVD_*(\struct C)=\rankVD_*(\struct B)\leq\alpha\oplus\beta$. Similarly, ${\rankVD_*(\struct C)\leq\alpha\oplus\beta}$ if $\beta=0$.

\subparagraph*{Inductive step.} $\alpha>0$ and $\beta>0$.\\
 If $\struct A$ is a finite sum of elements from $\VD_{<\alpha}$, then $\rankVD_*(\struct A)<\alpha$ and $\rankVD_*(\struct C)<\alpha\oplus\beta$ by induction. Similarly, $\rankVD_*(\struct C)<\alpha\oplus\beta$ if $\struct B$ is a finite sum. It remains to show the claim under the assumption that $\struct A$ and $\struct B$ are $\omega$\nobreakdash-sums or $\omega^*$-sums. We distinguish four cases. In each case, let $c_1\colon A^2\to Q_1$ and $c_1\colon B^2\to Q_2$ be $\leq$-colourings of $\struct A$ and $\struct B$ such that
\begin{equation*}
	c\colon(A\times B)^2\to Q_1\times Q_2,
	\bigl((a_1,b_1),(a_2,b_2)\bigr)\mapsto \bigl(c_1(a_1,a_2),c_2(b_1,b_2)\bigr)
\end{equation*}
is a $\leq$-colouring of $\struct C$. 

\subparagraph*{Case 1.} $\struct A$ is an $\omega$-sum of elements from $\VD_{<\alpha}$ and $\struct B$ is an $\omega^*$-sum of elements from~$\VD_{<\beta}$.\\
By Lemma~\ref{lemma:ramsey_sequence}, there exist a strictly increasing, unbounded sequence $(a_i)_{i\in\Nat}$ in $\struct A$ and a colour $q_1\in Q_1$ such that $c_1(a_i,a_j)=q_1$ for all $i,j\in\Nat$ with $i<j$. By the dual of Lemma~\ref{lemma:ramsey_sequence}, there exist a strictly decreasing, unbounded sequence $(b_i)_{i\in\Nat}$ in $\struct B$ and a colour $q_2\in Q_2$ such that $c_2(b_i,b_j)=q_1$ for all $i,j\in\Nat$ with $i>j$. Depending on how $(a_0,b_0)$ compares to $(a_1,b_1)$ in $\struct C$, we distinguish two cases.

\subparagraph*{Case 1.1.} $(a_0,b_0)<(a_1,b_1)$.\\
Figure~\ref{fig:case_1.1} depicts the idea behind the treatment of this case. The horizontal axis describes $\struct A$ and increases from left to right, wheres the vertical axis outlines $\struct B$ and grows from bottom to top. Within the grid, arrows point from smaller to greater elements.

\tikzstyle{point}=[circle,draw,fill,minimum size=0.7mm,inner sep=0]
\tikzstyle{pointname}=[scale=0.8,inner sep=0.5mm,outer sep=0,fill=white]
\tikzstyle{cmp}=[semithick,-latex']

\begin{figure}
\centering
\begin{tikzpicture}[semithick,scale=1.5]

\path[use as bounding box] (-7mm,7mm) rectangle (51mm,-51mm);

\draw[->] (0mm,0mm) -- +(50mm,0mm);
\draw[->] (0mm,-50mm) -- +(0mm,50mm);

\draw (22mm,2mm) node[anchor=south] {$\struct A$};
\draw (-1mm,-25mm) node[anchor=east] {$\struct B$};

\foreach \x/\i in {6mm/0,12mm/1,18mm/2,27mm/k-1,33mm/k,38mm/k+1,46mm/k+2}
{
	\draw (\x,-0.7mm) -- ++(0mm,1.4mm) +(0mm,1mm) node[anchor=base] {$a_{\scriptscriptstyle \i}$};
	\draw[dashed,thin] (\x,-6mm) -- +(0mm,-44mm);
}

\foreach \y/\j in {6mm/0,12mm/1,45mm/\ell}
{
	\draw (0.7mm,-\y) -- +(-1.4mm,0mm) node[anchor=east,inner sep=0.5mm] {$b_{\scriptscriptstyle \j}$};
}

\draw[dashed,thin] (0mm,-6mm) -- +(49mm,0);

\node[point,label={[pointname]above right:$(a_{\scriptscriptstyle 0},b_{\scriptscriptstyle 0})$}] (a0b0) at (6mm,-6mm) {};
\node[point,label={[pointname]above right:$(a_{\scriptscriptstyle 1},b_{\scriptscriptstyle 1})$}] (a1b1) at (12mm,-12mm) {};
\node[point,label={[pointname]below:$(a,b)$}] (ab) at (30mm,-20mm) {};
\node[point,label={[pointname]above right:$(a_{\scriptscriptstyle k},b_{\scriptscriptstyle 0})$}] (akb0) at (33mm,-6mm) {};
\node[point,label={[pointname]below right:$(a_{\scriptscriptstyle k+1},b_{\scriptscriptstyle \ell})$}] (ak1bl) at (38mm,-45mm) {};
\node[point,label={[pointname]above:$(a',b')$}] (a'b') at (42mm,-40mm) {};
\draw[cmp] (a0b0) edge (a1b1)
(ab) edge (akb0) (akb0) edge (ak1bl) (ak1bl) edge (a'b');

\draw (22.5mm,-3mm) node {$Y_{\scriptscriptstyle 1}$};
\draw (3mm,-30mm) node[anchor=base] {$X_{\scriptscriptstyle 0}$};
\draw (9mm,-30mm) node[anchor=base] {$X_{\scriptscriptstyle 1}$};
\draw (15mm,-30mm) node[anchor=base] {$X_{\scriptscriptstyle 2}$};
\draw (22.5mm,-30mm) node[anchor=base] {$\dotsb$};
\draw (30mm,-30mm) node[anchor=base] {$X_{\scriptscriptstyle k}$};
\draw (42mm,-30mm) node[anchor=base] {$X_{\scriptscriptstyle k+2}$};
\end{tikzpicture}
\caption{Proof sketch for Case~1.1.}
\label{fig:case_1.1}
\end{figure}
Formally, let
\begin{align*}
	X_0 &= (-\infty,a_0]_{\struct A}\times(-\infty,b_0)_{\struct B} &
	X_k &= (a_{k-1},a_k]_{\struct A}\times(-\infty,b_0)_{\struct B}\ \text{for $k\geq 1$}
\end{align*}
and
\begin{align*}
	Y_1 &= A\times[b_0,\infty)_{\struct B} &
	Y_2 &= \bigcup_{k\in\Nat} X_{2k} &
	Y_3 &= \bigcup_{k\in\Nat} X_{2k+1}\,.
\end{align*}
Since ${A\times B=Y_1\uplus Y_2\uplus Y_3}$, by Proposition~\ref{prop:VD_sum}, it suffices to show $\rankVD_*(Y_i)\leq\alpha\oplus\beta$ for $i=1,2,3$. Lemma~\ref{lemma:sequence_split} and its dual yield
\begin{align*}
	\rankVD_*\bigl((-\infty,a_0]_{\struct A}\bigr)&<\alpha &
	\rankVD_*\bigl((a_{k-1},a_k]_{\struct A}\bigr)&<\alpha\ \text{for $k\geq 1$} &
	\rankVD_*\bigl([b_0,\infty)_{\struct B}\bigr)<\beta\,.
\end{align*}

\noindent Together with the induction hypothesis this yields $\rankVD_*(X_k)<\alpha\oplus\beta$ for all $k\in\Nat$ as well as $\rankVD_*(Y_1)<\alpha\oplus\beta$.

As a next step, we show that
\begin{equation}
\label{eq:stripe_order}
	X_k\ll X_{k+2}\ \text{for all $k\in\Nat$.}
\end{equation}
Therefore, let $(a,b)\in X_k$ and $(a',b')\in X_{k+2}$. Since the sequence of the $b_i$ is strictly decreasing and unbounded, there is an $\ell\geq1$ such that $b_\ell\leq b'$. The choice of the sequences $(a_i)_{i\in\Nat}$ and $(b_i)_{i\in\Nat}$ implies
\begin{equation*}
	c\bigl((a_0,b_0),(a_1,b_1)\bigr) = (q_1,q_2) = c\bigl((a_k,b_0),(a_{k+1},b_\ell)\bigr)
\end{equation*}
and hence $(a_k,b_0)<(a_{k+1},b_\ell)$. Since $\struct C$ is a linearisation of $\struct A\times\struct B$, we have $(a,b)<(a_k,b_0)$ and $(a_{k+1},b_\ell)<(a',b')$. Altogether,
\begin{equation*}
	(a,b) < (a_k,b_0) < (a_{k+1},b_\ell) < (a',b')\,.
\end{equation*}

\noindent As as a direct consequence of Eq.~\eqref{eq:stripe_order}, we obtain
\begin{align*}
	\struct A\restrict{Y_2} &= \sum_{k\in\omega} \struct A\restrict{X_{2k}} &
	\struct A\restrict{Y_3} &= \sum_{k\in\omega} \struct A\restrict{X_{2k+1}}\,.
\end{align*}
Since every $\struct A\restrict{X_{2k}}$ is a finite sum of elements from $\VD_{<\alpha\oplus\beta}$, $\struct A\restrict{Y_2}$ is an $\omega$-sum of elements from $\VD_{<\alpha\oplus\beta}$ and hence $\rankVD_*(Y_2)\leq\alpha\oplus\beta$. Analogously, $\rankVD_*(Y_3)\leq\alpha\oplus\beta$. This completes Case~1.1.

\subparagraph*{Case 1.2.} $(a_0,b_0)>(a_1,b_1)$.\\
This case is very similar to Case~1.1 and depicted in Figure~\ref{fig:case_1.2}. To see this, let
\begin{align*}
	X_0 &= (a_0,\infty)_{\struct A}\times[b_0,\infty)_{\struct B} &
	X_k &= (a_0,\infty)_{\struct A}\times[b_i,b_{i-1})_{\struct B}\ \text{for $k\geq 1$}
\end{align*}
and
\begin{align*}
	Y_1 &= (-\infty,a_0]_{\struct A}\times B &
	Y_2 &= \bigcup_{k\in\Nat} X_{2k} &
	Y_3 &= \bigcup_{k\in\Nat} X_{2k+1}\,.
\end{align*}
Again, we obtain $\rankVD_*(X_k)<\alpha\oplus\beta$ for all $k\in\Nat$ as well as $\rankVD_*(Y_1)<\alpha\oplus\beta$. Moreover, for each $k\in\Nat$ it holds that $X_k\gg X_{k+2}$ and hence
\begin{align*}
	\struct A\restrict{Y_2} &= \sum_{k\in\omega^*} \struct A\restrict{X_{2k}} &
	\struct A\restrict{Y_3} &= \sum_{k\in\omega^*} \struct A\restrict{X_{2k+1}}\,.
\end{align*}
Consequently, $\rankVD_*(Y_2),\rankVD_*(Y_3)\leq\alpha\oplus\beta$. This completes Case~1.2 and hence Case~1.

\begin{figure}
\centering
\begin{tikzpicture}[semithick,scale=1.5]

\path[use as bounding box] (-7mm,7mm) rectangle (51mm,-51mm);

\draw[->] (0mm,0mm) -- +(50mm,0mm);
\draw[->] (0mm,-50mm) -- +(0mm,50mm);

\draw (25mm,2mm) node[anchor=south] {$\struct A$};
\draw (-1mm,-22.5mm) node[anchor=east] {$\struct B$};

\foreach \x/\i in {6mm/0,12mm/1,45mm/\ell}
{
	\draw (\x,-0.7mm) -- ++(0mm,1.4mm) +(0mm,1mm) node[anchor=base] {$a_{\scriptscriptstyle \i}$};
}

\foreach \y/\j in {6mm/0,12mm/1,18mm/2,27mm/k-1,33mm/k,38mm/k+1,46mm/k+2}
{
	\draw (0.7mm,-\y) -- +(-1.4mm,0mm) node[anchor=east,inner sep=0.5mm] {$b_{\scriptscriptstyle \j}$};
	\draw[dashed,thin] (6mm,-\y) -- +(43mm,0mm);
}

\draw[dashed,thin] (6mm,0mm) -- +(0mm,-50mm);

\node[point,label={[pointname]above right:$(a_{\scriptscriptstyle 0},b_{\scriptscriptstyle 0})$}] (a0b0) at (6mm,-6mm) {};
\node[point,label={[pointname]above right:$(a_{\scriptscriptstyle 1},b_{\scriptscriptstyle 1})$}] (a1b1) at (12mm,-12mm) {};
\node[point,label={[pointname]above:$(a,b)$}] (ab) at (20mm,-30mm) {};
\node[point,label={[pointname,xshift=1mm]above left:$(a_{\scriptscriptstyle 0},b_{\scriptscriptstyle k})$}] (a0bk) at (6mm,-33mm) {};
\node[point,label={[pointname,yshift=1mm]above:$(a_{\scriptscriptstyle \ell},b_{\scriptscriptstyle k+1})$}] (albk1) at (45mm,-38mm) {};
\node[point,label={[pointname]below:$(a',b')$}] (a'b') at (40mm,-42mm) {};
\draw[cmp] (a1b1) edge (a0b0)
(a'b') edge (albk1) (albk1) edge (a0bk) (a0bk) edge (ab);

\draw (3mm,-22.5mm) node {$Y_{\scriptscriptstyle 1}$};
\draw (28mm,-3mm) node {$X_{\scriptscriptstyle 0}$};
\draw (28mm,-9mm) node {$X_{\scriptscriptstyle 1}$};
\draw (28mm,-15mm) node {$X_{\scriptscriptstyle 2}$};
\draw (28mm,-21.5mm) node {$\vdots$};
\draw (28mm,-30mm) node {$X_{\scriptscriptstyle k}$};
\draw (28mm,-42mm) node {$X_{\scriptscriptstyle k+2}$};
\end{tikzpicture}
\caption{Proof sketch for Case~1.2.}
\label{fig:case_1.2}
\end{figure}
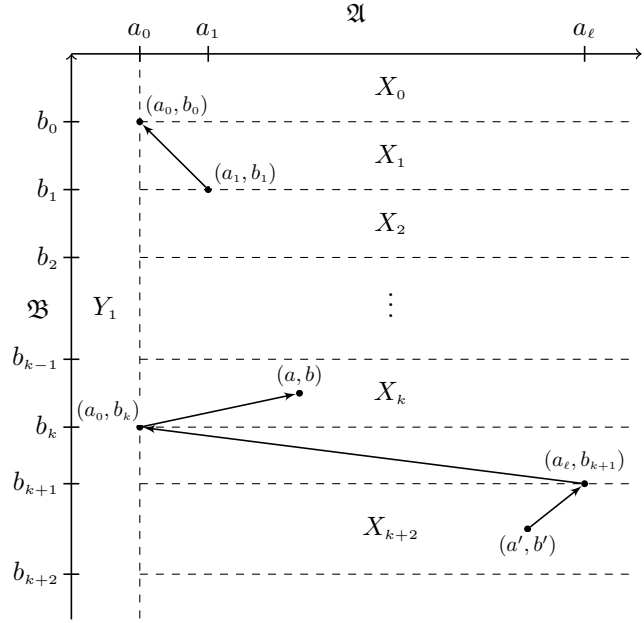

\subparagraph*{Case 2.} $\struct A$ and $\struct B$ both are $\omega$-sums.\\
Consider the strictly increasing, unbounded sequences $(a_i)_{i\in\Nat}$ in $\struct A$ and $(b_i)_{i\in\Nat}$ in $\struct B$ which exist by Lemma~\ref{lemma:ramsey_sequence}. Depending on how $(a_0,b_1)$ compares to $(a_1,b_0)$ in $\struct C$, we distinguish two cases.

\subparagraph*{Case 2.1.} $(a_0,b_1)<(a_1,b_0)$.\\
This case is treated similar to Case~1.1 and depicted in Figure~\ref{fig:case_2.1}.

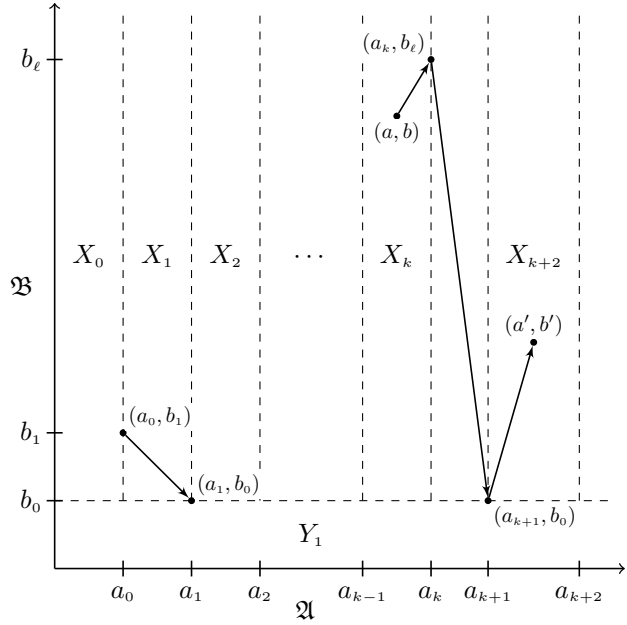
\begin{figure}
\centering
\begin{tikzpicture}[semithick,scale=1.5]

\path[use as bounding box] (-7mm,-7mm) rectangle (51mm,51mm);

\draw[->] (0mm,0mm) -- +(50mm,0mm);
\draw[->] (0mm,0mm) -- +(0mm,50mm);

\draw (22mm,-2mm) node[anchor=north] {$\struct A$};
\draw (-1mm,25mm) node[anchor=east] {$\struct B$};

\foreach \x/\i in {6mm/0,12mm/1,18mm/2,27mm/k-1,33mm/k,38mm/k+1,46mm/k+2}
{
	\draw (\x,0.7mm) -- ++(0mm,-1.4mm)  node[anchor=north] {$a_{\scriptscriptstyle \i}$};
	\draw[dashed,thin] (\x,6mm) -- +(0mm,43mm);
}

\foreach \y/\j in {6mm/0,12mm/1,45mm/\ell}
{
	\draw (0.7mm,\y) -- +(-1.4mm,0mm) node[anchor=east,inner sep=0.5mm] {$b_{\scriptscriptstyle \j}$};
}

\draw[dashed,thin] (0mm,6mm) -- +(49mm,0mm);

\node[point,label={[pointname]above right:$(a_{\scriptscriptstyle 0},b_{\scriptscriptstyle 1})$}] (a0b1) at (6mm,12mm) {};
\node[point,label={[pointname]above right:$(a_{\scriptscriptstyle 1},b_{\scriptscriptstyle 0})$}] (a1b0) at (12mm,6mm) {};
\node[point,label={[pointname]below:$(a,b)$}] (ab) at (30mm,40mm) {};
\node[point,label={[pointname]above left:$(a_{\scriptscriptstyle k},b_{\scriptscriptstyle \ell})$}] (akbl) at (33mm,45mm) {};
\node[point,label={[pointname]below right:$(a_{\scriptscriptstyle k+1},b_{\scriptscriptstyle 0})$}] (ak1b0) at (38mm,6mm) {};
\node[point,label={[pointname]above:$(a',b')$}] (a'b') at (42mm,20mm) {};
\draw[cmp] (a0b1) edge (a1b0)
(ab) edge (akbl) (akbl) edge (ak1b0) (ak1b0) edge (a'b');

\draw (22.5mm,3mm) node {$Y_{\scriptscriptstyle 1}$};
\draw (3mm,27mm) node[anchor=base] {$X_{\scriptscriptstyle 0}$};
\draw (9mm,27mm) node[anchor=base] {$X_{\scriptscriptstyle 1}$};
\draw (15mm,27mm) node[anchor=base] {$X_{\scriptscriptstyle 2}$};
\draw (22.5mm,27mm) node[anchor=base] {$\dotsb$};
\draw (30mm,27mm) node[anchor=base] {$X_{\scriptscriptstyle k}$};
\draw (42mm,27mm) node[anchor=base] {$X_{\scriptscriptstyle k+2}$};
\end{tikzpicture}
\caption{Proof sketch for Case~2.1.}
\label{fig:case_2.1}
\end{figure}

\subparagraph*{Case 2.2.} $(a_0,b_1)>(a_1,b_0)$.\\
This case is symmetric to Case~2.1.

\subparagraph*{Case 3.} $\struct A$ is an $\omega^*$-sum and $\struct B$ is an $\omega$-sum.\\
This case is symmetric to Case~1.

\subparagraph*{Case 4.} $\struct A$ and $\struct B$ both are $\omega^*$-sums.\\
This case is dual to Case~2.

\bigskip
\noindent This finishes the proof of Proposition~\ref{prop:VD_box2}.
\end{proof}

\noindent Finally, we are in a position to perform the induction which proves Proposition~\ref{prop:VD_box}.

\begin{proof}[Proof of Proposition~\ref{prop:VD_box}]
\label{proof:prop:VD_box}
We show the claim by induction on $n$.

\subparagraph*{Base case.} $n=1$.\\
Clearly, $\struct A\cong\struct B_1$ and hence $\rankVD_*(\struct A)=\rankVD_*(\struct B_1)$.

\subparagraph*{Inductive step.} $n>1$.\\
To simplify notation, we assume that $\struct A$ is a linearisation of $\struct B_1\times\dotsb\times\struct B_n$. For each $i$ let $c_i\colon B_i^2\to Q_i$ be a $\leq$-colouring of $\struct B_i$ such that
\begin{equation*}
	c\colon(B_1\times\dotsb\times B_n)^2\to Q_1\times\dotsb\times Q_n,
	(\bar a,\bar b)\mapsto\bigl(c_1(a_1,b_1),\dotsc,c_n(a_n,b_n)\bigr)
\end{equation*}
is a $\leq$-colouring of $\struct A$. We consider the relation $\sim$ on $B_1$ which is defined by $x\sim y$ iff $c_1(x,x)=c_1(y,y)$. This is an equivalence relation with at most $|Q_1|$ equivalence classes, say $X_1,\dotsc,X_m\subseteq B_1$ are these $\sim$-classes. Obviously, $\struct A$ is a sum augmentation of the $m$ orderings $\struct A\restrict{(X_i\times B_2\times\dotsb\times B_n)}$ for $\run i1m$. By Proposition~\ref{prop:VD_sum}, it suffices to show for each $i$ the inequality
\begin{equation}
	\label{eq:VD_prodn}
	\rankVD_*\bigl(\struct A\restrict{(X_i\times B_2\times\dotsb\times B_n)}\bigr)\leq
	\rankVD_*(\struct B_1)\oplus\dotsb\oplus\rankVD_*(\struct B_n)\,.
\end{equation}

Therefore, define for each $x\in B_1$ a scattered linear ordering $\struct C_x$ by $\univ{\struct C_x}=B_2\times\dotsb\times B_n$ and $\bar a\leq^{\struct C_x}\bar b$ iff $(x,\bar a)\leq^{\struct A}(x,\bar b)$. Clearly, $\struct C_x$ is a tame box augmentation of $\struct B_2,\dotsc,\struct B_n$ and hence
\begin{equation}
	\label{eq:VD_prodn-1}
	\rankVD_*(\struct C_x)\leq
	\rankVD_*(\struct B_2)\oplus\dotsb\oplus\rankVD_*(\struct B_n)
\end{equation}
by induction. For $x,y\in B_1$ with $x\sim y$ and all $\bar a,\bar b\in B_2\times\dotsb\times B_n$ we have ${c\bigl((x,\bar a),(x,\bar b)\bigr)=c\bigl((y,\bar a),(y,\bar b)\bigr)}$ and hence $\bar a\leq^{\struct C_x} \bar b$ iff $\bar a\leq^{\struct C_y}\bar b$, i.e., $\struct C_x=\struct C_y$. For any $\sim$\nobreakdash-class $X_i\subseteq B_1$ and every $x\in X_i$ we obtain that $\struct A\restrict{(X_i\times B_2\times\dotsb\times B_n)}$ is a tame box augmentation of $\struct B_1\restrict{X_i}$ and $\struct C_x$. Finally, Eq.~\eqref{eq:VD_prodn} follows from $\rankVD_*(\struct B_1\restrict{X_i})\leq\rankVD_*(\struct B_1)$, Eq.~\eqref{eq:VD_prodn-1}, and Proposition~\ref{prop:VD_box2}.
\end{proof}

\subsection{Proof of the Main Result}

In order to conclude Theorem~\ref{thm:main} from Corollaries~\ref{cor:delhomme_indecomposability}, \ref{cor:VD_sum_indecomposable}, and \ref{cor:VD_box_indecomposable}, we need another auxiliary result. Statement (1) of the lemma below is in fact shown by the proof of Proposition~4.5 in~\cite{KRS05}.

\begin{lemma}
\label{lemma:scattered_closed_intervals}
Let $\struct A$ be a linear ordering and $\alpha<\rankFC(\struct A)$.
\begin{enumerate}
\item $\struct A$ contains a scattered closed interval $I$ with $\rankFC(I)=\alpha+1$.
\item $\struct A$ contains a scattered closed interval $I$ with $\rankVD_*(I)=\alpha$.
\end{enumerate}
\end{lemma}

\begin{proof}
We only show (2). By (1), there exists a closed scattered interval $I$ of $\struct A$ with $\rankVD(I)=\rankFC(I)=\alpha+1$. Since $I$ has a least and a greatest element, it is neither an $\omega$-sum nor an $\omega^*$-sum nor a $\zeta$-sum of elements from $\VD_{<\alpha+1}=\VD_\alpha$. Thus, $I$ is a finite sum of elements from $\VD_\alpha$ and hence $\rankVD_*(I)\leq\alpha$. Due to Eq.~\eqref{eq:VD_inequality}, $\rankVD_*(I)=\alpha$.
\end{proof}

\noindent Now, we are prepared to provide the missing proof of the main result.

\begin{proof}[Proof of Theorem~\ref{thm:main}]
By contradiction, assume there exists a tree-automatic linear ordering $\struct A$ with $\rankFC(\struct A)\geq\omega^\omega$. Consider the formula $\phi(x,y_1,y_2) = y_1\leq x\land x\leq y_2$. By Lemma~\ref{lemma:scattered_closed_intervals}, for each $d\in\Nat$ there exists a scattered closed interval $I=[b_1,b_2]_{\struct A}$ in $\struct A$ with $b_1\leq b_2$ and $\rankVD_*(I)=\omega^d$. Since $I=\phi^{\struct A}(\cdot,b_1,b_2)$ and $\omega^d$ is $\rankVD_*$-sum-indecomposable as well as $\rankVD_*$-tame-box-indecomposable, this contradicts Corollary~\ref{cor:delhomme_indecomposability}.
\end{proof}

\section{$\Tbin$-Free Tree-Automatic Presentations}

In this section, we investigate a restricted form of tree-automaticity where only those tree-automatic presentations $\bigl(\aut A;(\aut A_R)_{R\in\Rel}\bigr)$ are permitted for which the binary tree
\begin{equation*}
	T(\aut A) = T\bigl(L(\aut A)) = \bigcup_{t\in L(\aut A)} \dom(t)
\end{equation*}
is of bounded branching complexity---in some sense defined later.\footnote{Roughly speaking, the branching complexity is bounded if the infinite full binary tree cannot be embedded and is measured in terms of the Cantor-Bendixson rank.} The main result of this section, namely Theorem~\ref{thm:main_bounded_rank}, states that any linear ordering $\struct A$ which admits a tree-automatic presentation whose branching complexity is bounded by $k\in\Nat$ satisfies ${\rankFC(\struct A)<\omega^k}$.

\subsection{Binary Trees and the Cantor-Bendixson Rank}

The \emph{infinite full binary tree} is the set $\Tbin=\{0,1\}^\star$ whose nodes are ordered by the prefix-relation $\preceq$. A \emph{binary tree} is a (possibly empty) prefix-closed subset $T\subseteq\Tbin$. The (isomorphism type of) the \emph{subtree} rooted at $u\in T$ is
\begin{equation*}
	T\restrict u = \Set{ v\in\{0,1\}^\star | uv\in T }\,.
\end{equation*}
A binary tree $T$ is \emph{regular} if it is a regular language. Due to the Myhill-Nerode theorem, this is equivalent to the fact that $T$ has (up to isomorphism) only finitely many distinct subtrees $T\restrict u$. To every tree language $L\subseteq\trees\Sigma$ we assign a binary tree
\begin{equation*}
	T(L) = \bigcup_{t\in L} \dom(t)\,.
\end{equation*}

\begin{lemma}
For every regular tree language $L\subseteq\trees\Sigma$ the binary tree $T(L)$ is regular.
\end{lemma}

\begin{proof}
Let $\aut A$ be a tree automaton recognising $L$. For each $u\in T(L)$ let
\begin{equation*}
	Q(u) = \set{ \aut A(t,u) | t\in L }\,.
\end{equation*}
It is easy to see that $Q(u)=Q(v)$ implies $T(L)\restrict u=T(L)\restrict v$. Thus, $T(L)$ is regular.
\end{proof}

\noindent A binary tree $T$ is called \emph{$\Tbin$-free} if $\Tbin$ cannot be embedded into $T$, i.e., there is no injection $f\colon \Tbin\to T$ such that $u\preceq v$ iff $f(u)\preceq f(v)$ for all $u,v\in\Tbin$. An \emph{infinite branch} of a binary tree $T$ is an infinite subset $P\subseteq T$ which is prefix-closed and linearly ordered by~$\preceq$. The \emph{derivative} of $T$ is the set $d(T)$ of all $u\in T$ which are contained in at least two distinct infinite branches of $T$. Clearly, $d(T)$ is a binary tree. For $n\in\Nat$ let $d^{(n)}(T)$ be the $n^{\mathrm{th}}$ derivation of $T$, i.e., $d^{(0)}(T)=T$ and $d^{(n)}(T)=d\bigl(d^{(n-1)}(T)\bigr)$ for $n>0$. Whenever $T$ is regular there exists an $n\in\Nat$ such that $d^{(n)}(T)=d^{(k)}(T)$ for all $k\geq n$ and $d^{(n)}(T)$ is finite precisely if $T$ is $\Tbin$-free~\cite{KRS05}. 

\begin{definition}
Let $T$ be a regular, $\Tbin$-free binary tree. The \emph{$\rankCB_*$-rank} of $T$, denoted by $\rankCB_*(T)$, is the least $n\in\Nat$ such that $d^{(n)}(T)$ is finite.\footnote{In fact, $\rankCB_*$ is a variation of the Cantor-Bendixson rank which was adapted to trees in \cite{KRS05}.}
\end{definition}

\noindent Clearly, $d(T\restrict u) = d(T)\restrict u$ and hence $\rankCB_*(T\restrict u)\leq\rankCB_*(T)$ for all $u\in T$.

\begin{definition}
A tree-automatic presentation $\bigl(\aut A;(\aut A_R)_{R\in\Rel}\bigr)$ is \emph{$\Tbin$-free} if $T\bigl(L(\aut A)\bigr)$ is $\Tbin$-free and then its \emph{rank} is the $\rankCB_*$-rank of $T\bigl(L(\aut A)\bigr)$.\footnote{In \cite{BGR11} the authors speak of \emph{bounded-rank tree-automatic presentations}. Their notion of \emph{rank} is defined differently, but can be shown to be equivalent to ours.}
\end{definition}

\begin{remark}
Obviously, the structures which admit a $\Tbin$-free tree-automatic presentation of rank $0$ are precisely the finite structures. Furthermore, it can be shown that the structures which admit a presentation of rank at most $1$ are exactly the string-automatic structures.\footnote{String-automatic structures are defined like tree-automatic structures but with finite words and finite automata instead of trees and tree automata.}
\end{remark}

\subsection{$\Tbin$-Free Tree-Automatic Presentation of Linear Orderings}

The following is the main result of this section.

\begin{theorem}
\label{thm:main_bounded_rank}
Let $\struct A$ be a linear ordering which admits a $\Tbin$-free tree-automatic presentation of rank $k\geq 1$. Then
\begin{equation*}
	\rankFC(\struct A) < \omega^k\,.
\end{equation*}
\end{theorem}

\begin{corollary}
\label{cor:main_ordinal_bounded_rank}
An ordinal $\alpha$ admits a $\Tbin$-free tree-automatic presentation of rank at most $k$ if, and only if,
\begin{equation*}
	\alpha < \omega^{\omega^k}\,.
\end{equation*}
\end{corollary}

\begin{remark}
As direct consequence of this corollary and Corollary~\ref{cor:main_ordinal}, every tree-automatic ordinal already admits a $\Tbin$-free tree-automatic presentation. In fact, Jain, Khoussainov, Schlicht, and Stephan~\cite{JKS12} recently showed that every tree-automatic presentation of an ordinal---or more generally, of a scattered linear ordering---is $\Tbin$-free.
\end{remark}

\noindent The proof of Theorem~\ref{thm:main_bounded_rank} works by more detailed inspection of the proofs of Theorem~\ref{thm:delhomme}, Corollary~\ref{cor:delhomme_indecomposability}, and Theorem~\ref{thm:main} in combination with the following lemma.

\begin{lemma}
\label{lemma:anti_chains_bounded}
Let $T$ be a regular, $\Tbin$-free binary tree. Then there exists a constant $C\in\Nat$ such that any anti-chain $A\subseteq T$ contains at most $C$ elements $u$ with $\rankCB_*(T\restrict u)=\rankCB_*(T)$.
\end{lemma}

\begin{proof}
If $\rankCB_*(T)=0$ then $T$ is finite and the claim is trivially satisfied. Thus, assume $\rankCB_*(T)=k>0$. Let $n\in\Nat$ be the \emph{index} of $T$, i.e., the size of the set $\set{ T\restrict u | u \in T }$. We show that $C=2^n$ is a possible choice.

By contradiction, suppose there is an anti-chain $A$ consisting of $2^n+1$ elements $u\in T$ satisfying ${\rankCB_*(T\restrict u)=k}$. Let $B$ be the set of all $v\in T$ which are the longest common prefix of two distinct elements from $A$. Then $B$ contains exactly $2^n$ elements. For every $u\in A$ the set $d^{(k-1)}(T\restrict u)=d^{(k-1)}(T)\restrict u$ is infinite. By K\"onig's lemma, there exists an infinite branch of $d^{(k-1)}(T)$ containing $u$. Thus, $B\subseteq d^{(k)}(T)$. For every $v\in d^{(k)}(T)$ it holds that $d^{(k)}(T)\restrict v=d^{(k)}(T\restrict v)$ and hence the index of $d^{(k)}(T)$ is at most $n$. Since $d^{(k)}(T)$ contains at least $2^n$ elements, a simple pumping argument shows that $d^{(k)}(T)$ is infinite. But this contradicts $\rankCB_*(T)=k$.
\end{proof}

\noindent Now, we are in a position to show the main result of this section.

\begin{proof}[Proof of Theorem~\ref{thm:main_bounded_rank}]
We show the claim by induction on $k\geq 1$. Therein, we use the induction hypothesis only in the following restricted form: Every scattered linear ordering $\struct A$ which admits a $\Tbin$-free tree-automatic presentation of rank $k\geq0$ satisfies $\rankVD_*(\struct A)<\omega^k$. For $k\geq 1$ this assertion easily follows from $\rankVD(\struct A)=\rankFC(\struct A)<\omega^k$.

\subparagraph*{Base case.} $k=0$.\\
Since any structure which admits a $\Tbin$-free tree-automatic presentation of rank $0$ is finite, every such scattered linear ordering $\struct A$ trivially satisfies $\rankVD_*(\struct A)=0<\omega^0$.

\subparagraph*{Inductive step.} $k\geq 1$.\\
By contradiction, assume there exists a tree-automatic linear ordering $\struct A$ which admits a $\Tbin$-free tree-automatic presentation $\bigl(\aut A;(\aut A_R)_{R\in\Rel}\bigr)$ of rank $k$ and satisfies $\rankFC(\struct A)\geq\omega^k$. To keep notation simple, we assume that the naming function $\mu\colon A\to L(\aut A)$ is the identity, i.e., $\struct A$ is identified with its tree-automatic copy $\mu(\aut A)$. Let $C$ be the constant which exists by Lemma~\ref{lemma:anti_chains_bounded} for the binary tree $T(A)$. Moreover, let $\class S_\phi^{\struct A}$ be the set which is constructed in the proof of Theorem~\ref{thm:delhomme} from $\bar{\aut A}$ and the formula ${\phi(x,y_1,y_2) = y_1\leq x\land x\leq y_2}$.  We show that $\class S_\phi^{\struct A}$ contains for each $n\in\Nat$ a scattered linear ordering $\struct B$ with $\omega^{k-1} n<\rankVD_*(\struct B)<\omega^k$. This contradicts the finiteness of $\class S_\phi^{\struct A}$ and proves the theorem.

Therefore, consider some $n\in\Nat$. By Lemma~\ref{lemma:scattered_closed_intervals}, there exists a scattered closed interval ${I=[a_1,a_2]_{\struct A}}$ of $\struct A$ with $a_1\leq a_2$ and $\rankVD_*(I)=\omega^{k-1}(nC+1)$. Now, we delve into the details of the proof of Theorem~\ref{thm:delhomme}. Since $I=\phi^{\struct A}(\cdot,a_1,a_2)$ and $\omega^{k-1}(nC+1)$ is $\rankVD_*$-sum-indecomposable, there exists a $\sim_{(a_1,a_2)}$-class $B\subseteq I$ such that $\rankVD_*(B)=\omega^{k-1}(nC+1)$. Let $\vartheta=\bigl(t_D,U,(\rho_R)_{R\in\{\phi\}\uplus\Rel}\bigr)$ be the corresponding $(a_1,a_2)$-type, $u_1,\dotsc,u_r$ an enumeration of $U$, and $\struct S_i=\struct S_{\gamma(\vartheta,u_i)}$ for each $\run i1r$. Notice that the $\struct S_i$ are scattered linear orderings and form a tame box decomposition of $\struct A\restrict B$. It is easy to see that $T(S_i)\subseteq T(A)\restrict{u_i}$ and hence $\rankCB_*\bigl(T(S_i)\bigr)\leq k$ for each $i$. Since $U$ is an anti-chain in $T(A)$, equality holds true in at most $C$ cases. Without loss of generality, there exists a $p\leq C$ such that $\rankCB_*\bigl(T(S_i)\bigr)=k$ for $i\leq p$ and $\rankCB_*\bigl(T(S_i)\bigr)<k$ for $i>p$.

By the restricted induction hypothesis, we obtain $\rankVD_*(\struct S_i)<\omega^{k-1}$ for $i>p$. If we had $\rankVD_*(\struct S_i)\leq\omega^{k-1}n$ for each $\run i1p$, then
\begin{equation*}
	\underbrace{\rankVD_*(\struct S_1)\oplus\dotsb\oplus\rankVD_*(\struct S_p)}_{\leq\omega^{k-1}np}\oplus
		\underbrace{\rankVD_*(\struct S_{p-1})\oplus\dotsb\oplus\rankVD_*(\struct S_r)}_{<\omega^{k-1}} <
	\omega^{k-1}(nC+1)\,.
\end{equation*}
This would contradict Proposition~\ref{prop:VD_box} and hence there exists a $\elem j1p$ with ${\rankVD_*(\struct S_j)>\omega^{k-1}n}$. Since $\struct S_j$ can be embedded into $\struct A\restrict B$, we further obtain
\begin{equation*}
	\rankVD_*(\struct S_j)\leq\omega^{k-1}(nC+1)<\omega^k\,.\qedhere
\end{equation*}
\end{proof}

\noindent In order to verify Corollary~\ref{cor:main_ordinal_bounded_rank} we still have to prove that every ordinal $\alpha<\omega^{\omega^k}$ admits a $\Tbin$-free tree-automatic presentation of rank at most $k$.

\begin{proof}[Proof of Corollary~\ref{cor:main_ordinal_bounded_rank}]
The ``only if''-part follows directly from Theorem~\ref{thm:main_bounded_rank} and we only need to show the ``if''-part. For $k=0$ the claim is trivial since each ordinal $\alpha<\omega$ is finite. Thus, assume $k>0$ and consider some $\alpha<\omega^{\omega^k}$. There exists an $n\in\Nat$ such that $\alpha<\omega^{\omega^{k-1}n}$. The ordinal $\omega^{\omega^{k-1} n}$ can be regarded as the lexicographically ordered set of all $n$-tuples of elements from $\omega^{\omega^{k-1}}$. Let $\bar{\aut A}$ be the tree-automatic presentation of $\omega^{\omega^{k-1}}$ which was constructed in Lemma~\ref{lemma:ordinals_are_tree_automatic} and $\nu\colon A\to\trees\Sigma$ the corresponding naming function. A closer look at the induction in the proof of Lemma~\ref{lemma:ordinals_are_tree_automatic} reveals that $\bar{\aut A}$ is $\Tbin$-free and of rank $k$. The map $\mu\colon\omega^{\omega^{k-1}n}\to\trees{\Sigma_\Box^n}$ with
\begin{equation*}
	\mu(\beta_1,\dotsc,\beta_n)=\otimes\bigl(\nu(\beta_1),\dotsc,\nu(\beta_n)\bigr)
\end{equation*}
can be used as naming function for a $\Tbin$-free tree-automatic presentation of rank $k$ of $\omega^{\omega^{k-1}n}$. Finally, $\alpha$ is $\FO$-definable with one parameter in $\omega^{\omega^{k-1}n}$ and hence admits a $\Tbin$-free tree-automatic presentation of rank $k$ as well.
\end{proof}

\bibliographystyle{abbrv}
\bibliography{library}

\end{document}